\documentclass{rcdj}

\begin{document}

\setcounter{page}{91}
\journal{REGULAR AND CHAOTIC DYNAMICS, V.\,9, \No2, 2004}
\runningtitle{BILLIARDS, INVARIANT MEASURES, AND EQUILIBRIUM THERMODYNAMICS. II}
\title{BILLIARDS, INVARIANT MEASURES, AND~EQUILIBRIUM THERMODYNAMICS. II}
\runningauthor{V.\,V.\,KOZLOV}
\authors{V.\,V.\,KOZLOV}
{Steklov Mathematical Institute\\
Russian Academy of Sciences\\
117966 Moscow, Russia\\
E-mail: kozlov@pran.ru}

\abstract{The kinetics of collisionless continuous medium is studied in a
bounded region on a curved manifold. We have assumed that in statistical
equilibrium, the probability distribution density depends only on the
total energy. It is shown that in this case, all the fundamental relations
for a multi-dimensional ideal gas in thermal equilibrium hold true.}
\amsmsc{82C22, 70F07, 70F45}
\doi{10.1070/RD2004v009n02ABEH000268}
\received 03.11.2003.

\maketitle

According to Gibbs, the basic object of statistical mechanics is an
ensemble of identical Hamiltonian systems. The systems do not interact
with each other and the assembly of them makes essentially a collisionless
continuous medium. From the viewpoint of kinetics, the Hamiltonian systems
with elastic impacts, i.\,e. {\it billiards\/}, are especially
interesting. These are systems where particles move inertially inside a
bounded region and bounce elastically against the boundaries of the
region. As it is shown in Refs.~[1,\,2], in a billiards, the probability
distribution density as a function of time~$t$ (this function satisfies
the classic Liouville equation) necessarily has the {\it weak} limit
as~$t\rightarrow\pm\infty$. This result justifies the {\it Zeroth} Law of
Thermodynamics in the Gibbs theory. The weak limit is a first integral
of the Hamilton equations and depends, in the ergodic case, only on the
system's energy. It is noted in Ref.~[3] that this is very often justified
even without any ergodic hypothesis$\colon$ it is important here to keep
in mind the function class, to which the probability distribution density
function belongs.

In Ref.~[3], we developed the thermodynamics of billiards in the Euclidean
space. It turns out to be possible to extend these observations to the
general case of a curved configurational space.

Let~$M^n$ be a compact configurational space of a natural mechanical
system with~$n$ degrees of freedom,~$x=(x_1,\ldots,x_n)$ be  local
coordinates on~$M$, and~$y=(y_1,\ldots,y_n)$ be the conjugate canonical
momenta. The motion to be considered is inertial. So the Hamiltonian is a
positively defined quadratic form with respect to the momenta$\colon$
\begin{equation}
H=\frac{1}{2}\suml_{i,j=1}^na_{ij}(x)y_iy_j.
\end{equation}
If we denote the  matrix of coefficients~$||a_{i,j}||$ as~$A$, then~$H=
(Ay,y)/2$.

Let~$f(\cdot)$ be a nonnegative summable function of one variable. By
analogy with the Gibbs canonical distribution, we introduce the density of
the stationary probability distribution in the phase
space~$\Gamma=T^*M\colon$
\begin{equation}
\rho(x,y)=\frac{f(\bt H)}{\intl_{\mR^n}\intl_M f(\bt H)\,d^ny\,d^nx}.
\end{equation}
Here,the factor~$\bt$ is  introduced  to non-dimensionalize  the argument
of~$f$. It is customary to take~$\bt=1/k\tau$,
where~$k$ is the Boltzmann constant, and~$\tau$ is the absolute
temperature.

The denominator in~(2) is referred to in Ref.~[3] as the {\it generalized}
statistic integral. It can be easily expressed in terms of~$\tau$, the only
external thermodynamical parameter here being~$M$, Riemann's volume of the
manifold. To this effect, we perform the linear change of
variables~$y\mapsto p$:
$$
y=C(x)p,\quad C^TAC=E.
$$
Thus,
$$
F=\intl_{\mR^n}\intl_M f(\bt
H)\,d^ny\,d^nx=\intl_M\intl_{\mR^n}f\left(\frac{\bt}{2}\sum p_i^2\right)
(\det A)^{-\frac{1}{2}}\,d^nx\,d^np=\frac{bv}{(\sqrt{\bt})^n},
$$
where
$$
v=\intl_M(\det A^{-1})^{\frac{1}{2}}\,d^nx
$$
is the volume~$M$ with respect to Riemannian metric~(1),
$$
b=\frac{2\pi^{\frac{n}{2}}}{\Gamma\left(\frac{n}
{2}\right)}\intl_0^\infty r^{n-1}f\left(\frac{r^2}{2}\right)\,dr=\const,
$$
where~$\Gamma$ is the Euler gamma function.

Now we calculate the average kinetic energy$\colon$
\begin{gather*}
E=\frac{1}{F}\intl_{\mR^n}\intl_M\frac{1}{2}(Ay,y)f\left(\frac{\bt}{2}
(Ay,y)\right)\,d^nx\,d^ny=\\
=\frac{1}{F}\intl_{\mR^n}\intl_M\frac{1}{2}\sum p_j^2f\left(\frac{\bt}{2}
\sum p_j^2\right)(\det A)^{-\frac{1}{2}}\,d^nx\,d^np=\\
=\frac{a}{\bt b},
\end{gather*}
where
$$
a=\frac{\pi^{\frac{n}{2}}}{\Gamma\left(\frac{n}{2}\right)}
\intl_0^\infty r^{n+1}f\left(\frac{r^2}{2}\right)\,dr=\const.
$$
It is interesting to note that the average internal energy of a
collisionless medium does not depend on the volume, which correlates with
\emph{Joule's law} for ideal gas.

As it is shown in Ref.~[4], if a Hamiltonian is a homogeneous function  with
respect to momenta, then the quantities calculated using the general routines of
statistical mechanics and density~(2) satisfy the First and the Second
Laws of Thermodynamics. Let us calculate, for example, the thermodynamic
entropy. For this, we should first (according to Ref.~[3]) write down the
following relation:
$$
E=\vk\frac{\partial F} {\partial\bt},
$$
which gives the coefficient~$\vk$. According to the general theory, this
coefficient must be a function of the statistical integral~$F$. In the case in
question,
$$
\vk=-\frac{2a}{bn}\frac{1}{F}.
$$
Let~$\Phi(F)$ be the antiderivative of~$\vk(F)$. Then, as shown in
Ref.~[3], the thermodynamic entropy is given by the equation
$$
S=\bt\frac{\partial \Phi}{\partial\bt}-\Phi.
$$
Hence,
\begin{equation}
S=\frac{a}{b}+\frac{2a}{bn}\ln F=\const+\frac{2a}{bn}\left(\ln v+\frac{n}
{2}\ln\tau\right).
\end{equation}

In the case of the Gibbs canonical distribution (~$f(z)=e^{-z}$), one can easily show, upon integration by parts,  that $2a=nb$.

Usually, the entropy of ideal gas in a \emph{three-dimensional}
vessel~$\Pi$ with volume~$w$ is
\begin{equation}
N\ln w+\frac{3N}{2}\ln\tau+\const,
\end{equation}
where~$N$ is the number of gas particles (this expression is sometimes
multiplied by the Boltzmann constant~$k$, but we do without it). To
compare~(3) and~(4), let us consider the Boltzmann--Gibbs gas consisting
of~$N$ identical small balls, moving in a  vessel~$\Pi$. The balls
collide elastically with each other and with the walls of the vessel.
Then, obviously,~$n=3N$, while the volume~$v$ is approximately equal
to~$w^N$ (for, in the case of non-interacting balls, the configurational
space~$M$ of the system is the direct product of~$N$ copies of~$\Pi$).
Having made these remarks, we see that~(3) and~(4) become identical up to
the insignificant constant factor~$2a/nb$ that depends on the type of the
function~$f(\cdot)$ and on the number of  degrees of freedom in the
system.

On the other hand, the entropy in statistical mechanics is given by the
integral
\begin{equation}
S=-\intl_\Gamma\rho\ln\rho\,d^nx\,d^ny.
\end{equation}
In the case of the canonical distribution, this integral coincides with
the thermodynamical entropy. Of course, for more general distributions of the
form~(2), this remarkable Gibbs' result is not valid. However,
the Gibbs entropy~(5) looks as follows:
\begin{equation}
\frac{\gam}{b}+\ln F,
\end{equation}
where
$$
\gam=-\frac{2\pi^{\frac{n}{2}}}{\Gamma\left(\frac{n}{2}\right)}
\intl_0^\infty r^{n-1}f\left(\frac{r^2}{2}\right)\ln f\left(\frac{r^2}{2}
\right)\,dr=\const.
$$
We see that~(3) and~(6) coincide up to an insignificant additive constant
and a somewhat less insignificant constant positive factor.

The latter remark is very important for the kinetics of a collisionless
medium, especially for the validation of the Second Law in the case of
irreversible processes. The matter is that (as proved in Refs.~[1]
and~[5]) if we replace the density~$\rho$ in~(5) with its weak limit,
then the Gibbs entropy gets a nonnegative increment. If the entropies
from~(3) and~(5) were not so closely related, this general result would
not allow a natural thermodynamical interpretation.

In the case of {\it ergodic} billiards, we can make not only general
conclusions on increase in entropy in irreversible processes, but we can
also calculate these increments. As a simple example, let us consider the
case where a collisionless medium is initially enclosed in the
portion~$M_-\subset M$ (regions~$M_-$ and~$M\backslash M_-$ are separated
with a wall), being in statistical equilibrium. After removal of the wall,
the medium expands irreversibly, tending to fill the whole region~$M$.
During this, its internal energy (and, consequently, its temperature) does
not change. According to~(3), the entropy gets a positive increment,
proportional to logarithm of the  ratio of volumes $M_-/ M_+$, where~$M_+=M$. This
result is in a good agreement  with the predictions of the phenomenological
thermodynamics. We should, probably, also mention that ergodicity of the
Boltzmann--Gibbs gas for a vessel shaped as a rectangular parallelepiped
was ascertained by Ya.\,G.\,Sinai~[6].

According to Ref.~[3], the thermodynamical variable~$P$, conjugate to the
volume~$v$, is given as
\begin{equation}
P=-\frac{1}{\bt}\frac{\d\Phi}{\d{v}}.
\end{equation}

Hence,
\begin{equation}
P=\frac{2a}{nb}\frac{k\tau}{v}.
\end{equation}

This is the equation of state for the considered system in statistical
equilibrium. This equation is identical in form  with the classical Clapeyron
equation. If~$f(z)=e^{-z}$, then~$2a=nb$, and~(8) exactly fits the
Clapeyron equation for a mole of ideal gas. The physical meaning of the
variable~$P$ is pressure.

Let us return to the Boltzmann--Gibbs gas of~$N$ small balls in a
three-dimensional vessel with volume~$w$. Then~$n=3N$, and we can assume
that~$v=w^N$. Substituting this  expression into~(8), we obtain an
equation, which is different from the Clapeyron equation. However, there
is no contradiction here, for~$P$ stands for the pressure
of~$3N$-dimensional gas.  The pressure~$p$ in ordinary gas, as a thermodynamical
quantity, conjugate to the volume~$w$, is given by~(7), only~$\Phi$ should
first be presented as a function of~$\tau$ and~$w$$\colon$
$$
p=-\frac{1}{\bt}\frac{\d\Phi}{\d{w}}=-\frac{1}{\bt}\frac{\d\Phi}{\d{v}}\frac{d{v}}{d{w}}=
\frac{2a}{3b}\frac{k\tau}{w}.
$$
For the Maxwell distribution (where~$f(z)=e^{-z}$),~$2a/3b=N$, and we
obtain the classical ideal gas equations$\colon$
\begin{equation}
E=\frac{3}{2}Nk\tau,\qquad pw=Nk\tau.
\end{equation}

For non-Maxwellian distributions, the value of~$2a/3b$, surely, differs
from~$N$. However, within a wide range of distributions, for large~$N$,
this value is approximately equal to~$N$$\colon$
\begin{equation}
\lim_{N\rightarrow\infty}\frac{2a(N)}{3Nb(N)}=1.
\end{equation}
For example, this range includes distributions with densities
\begin{equation}
f(\frac{r^2}{2})=g(r)e^{-r^2/2},
\end{equation}
where~$g(r)$ is an arbitrary non-negative polynomial in~$r$.

Indeed,
$$
\intl_0^\infty{r}^{n+\alpha-1}e^{-r^2/2}dz=\frac{1}{n+\alpha}\intl_0^\infty{r}^{n+\alpha+1}e^{-r^2/2}dz.
$$
Since~$\frac{n}{n+\alpha}\rightarrow1$ when~$\alpha$ is fixed, this results
in the limit relation~(10). Recall that functions of the form~(11) are
referred to as partial sums of {\it Gram-Charlier series\/}, and are
commonly used to approximate the distribution densities of arbitrary
random variables. In the case in question,  such
an approximation is possible due to  the well-known observation (traced as far
back as to Boltzmann) that in the major portion of a high-dimensional
space, any distribution is close to normal (strict formulations and
discussion can be found, for example, in Ref.~[7]). Since~$N$, as a rule,
is extremely large (of the order of~$10^{23}$) and not precisely known, we
can as well use the classical equations~(9) instead of~$E=(a/b)k\tau$
and~$pw=(2a/3b)k\tau$.

Thus, we have built a complete (nonequilibrium) theory of ideal gas within
the framework of Gibbs' general approach, using the concept of weak limits
of probability distributions and the result concerning the ergodic
behaviour of the Boltzmann--Gibbs gas. As distinct from Boltzmann's
approach, we do not use any additional assumptions (like the condition of
statistical independence of double collisions). The substantial difference
from Boltzmann's approach is that in our theory the gas reaches statistical
(thermal) equilibrium both as~$t\rightarrow+\infty$ and
as~$t\rightarrow-\infty$, these equilibriums being identical. This fact is
fully consistent with the invertibility property of the equations of motion.

Note also that by virtue of~(10), the entropy equation~(3) yields the
classical formula~(4) for monatomic ideal gas. Besides, the statistical
entropy~(5) coincides (up to an additive constant) with the thermodynamical
entropy~(3) as~$t\rightarrow\infty$.

\section*{A supplement. Particle distribution functions}

Let
\begin{equation}
\rho_N(x_1,\ldots,x_N,t)
\end{equation}
be the distribution density of the Boltzmann--Gibbs gas, which is a system
of~$N$ small identical balls enclosed in a rectangular box;~$x_j$ denotes the coordinates and momenta of the~$j$-th ball. The function~(12) satisfies
the Liouville equation and the initial condition~$\rho_N(x,0)$ at~$t=0$.
According to Bogolyubov (see Ref.~[8]), it is useful to
introduce~$s$-particle distribution functions~$\rho_s(x_1,\ldots,x_s,t)$,
averaging density~(12) over~$x_{s+1},\ldots,x_N$. The particle
distribution functions satisfy the infinite chain of ``hooked'' equations,
a so-called BBGKY (Bogolyubov, Born, Green, Kirkwood and Yvon) chain.
Under some {\it additional} assumptions (specifically, that of molecular
chaos ``in the past''), in the case of rarefied Boltzmann--Gibbs gas, one
derives the kinetic Boltzmann equation for one-particle distribution
function~$\rho_1$. These assumptions are not self-evident, do not follow
from the principles of Gibbs' statistical mechanics, and are to certain
extent similar to Boltzmann's assumption of statistical independence of
the balls' velocities before a double collision. Two important facts
follow from the Boltzmann equation$\colon$

\begin{itemize}
\item [1)] Boltzmann's entropy
$$-\int\rho_1\ln\rho_1d^6x_1$$
monotonously increases with time, and
\item [2)] as~$t\rightarrow+\infty$, the distribution~$\rho_1$ tends to
the Maxwell distribution.
\end{itemize}

However, these conclusions (at least, the former) cannot be
\emph{directly} verified by experiment. The matter is that the
thermodynamical entropy is introduced only for equilibrium states. The
ideas of determination of entropy for nonequilibrium states (like those
given, for example, in Refs.~[9,10]) are of methodical nature, proposing
to introduce an infinite number of additional internal thermodynamical
parameters. Using these ideas, the reader can find, for example, the entropy
of ideal gas as a function of time as the gas is adiabatically expanding
into vacuum (Joule's classical experiment).

We develop a different approach in the nonequilibrium statistical
mechanics of the Boltzmann--Gibbs gas. It has nothing to do with the
analysis of  {\it additional assumptions\/} that can be used  to close
Bogolyubov's chain of equations. We evolve Gibbs' classical principles and
try to avoid entirely additional assumptions of {\it conceptual} nature.
The crucial idea of our approach is$\colon$ transition to thermodynamical
(statistical) equilibrium is equal to replacement of the distribution~(12)
with its weak limit. This idea arises very naturally when one proceeds
from microscopic to macroscopic description of a dynamical system. The
weak limit (as~$t\rightarrow+\infty$ and~$t\rightarrow-\infty$) of
density~(12) (if it exists) coincides with  Birkhoff's average
$$
\bar{\rho}(x_1,\ldots,x_N).
$$

Let~$\varphi(x_1)$ be a test function. Then
$$
\begin{aligned}
\int\varphi\rho_1d^6x_1=\int\varphi\rho_Nd^6x_1\ldots{d}^6x_N
\xrightarrow{t\rightarrow\infty}\\
\longrightarrow\int\varphi\overline{\rho}_Nd^6x_1\ldots{d}^6x_N=\int\varphi\overline{\rho}_1d^6x_1,
\end{aligned}
$$
where
\begin{equation}
\overline{\rho}_1(x_1)=\int\overline{\rho}_Nd^6x_2\ldots{d}^6x_N.
\end{equation}
If the initial system with~$3N$ degrees of freedom is ergodic,
then~$\overline{\rho}_N$ is a summable function that depends only on the total
energy$\colon$
\begin{equation}
\overline{\rho}_N=f(\frac{\bt{m}}{2}(v_1^2+\ldots+v_N^2))/v\intl_{\mR^{3N}}fd^3v_1\ldots{d}^3v_N.
\end{equation}
Here~$m$ is the mass of the points, $v_j^2$ is the squared velocity of
the~$j$-th ball, the parameter~$\bt$ has the dimension  of the inverse of energy
(it is introduced to non-dimensionilize the argument of
~$f$) and~$v$ is the volume of the~$3N$-dimensional
configurational space of the system of~$N$ balls. The denominator in~(14)
makes the integral of~$\rho_N$ over the \emph{whole} phase space equal to
unity.

Formula~(14) shows that every possible position of the~$N$ balls is
equiprobable. This fact, noted earlier in Ref.~[2], means that a
\emph{homogeneous} distribution is established in the state of thermal
equilibrium. In fact, a similar conclusion follows Boltzmann's
theory$\colon$ the density of gas particles gets equalized and at the same
time the Maxwell velocity distribution is established.

Hence, the limit one-particle distribution~$\overline{\rho}_1$ does not
depend on the coordinates, and therefore, the averaging in~(13) can be
replaced with merely averaging over the velocities~$v_2,\ldots,v_N$. As a
result, the following simple expression is obtained$\colon$
\begin{equation}
\overline{\rho}_1(u)=\frac{\intl_{\mR^{3N-3}}f\left(\frac{1}{2\varkappa}
(u^2+v^2_2+\ldots+v^2_N)\right)d^3v_2\ldots{d}^3v_N}
{\intl_{\mR^{3N}}f\left(\frac{1}{2\varkappa}(v^2_1+v^2_2+\ldots+v^2_N)
\right)d^3v_1\ldots{d}^3v_N}
\end{equation}
where~$\varkappa=\bt/m,\,u\in\mR^3.$

It turns out that when certain additional constraints (of {\it analytical}
nature, not statistical) are imposed on the function~$f$, the limit
one-particle distribution function~$\overline{\rho}_1$ tends to the
Maxwell distribution as~$N\rightarrow\infty$. That is, for nearly {\it
any} initial distribution~$\rho_N(x_1,\ldots,x_N,0)$ (even without
assuming that~$\rho_N$ is symmetrical relative to~$x_1,\ldots,x_N$), the
balls' velocity distribution in the state of thermal equilibrium is, to
all practical purpose, normal (if, as usual,~$N$ is sufficiently large).

It should be underlined that, in such an approach, it is  meaningless  to
speak of the  rate of convergence of~$\rho_1$ (as a function of time) to the
limit distribution~$\overline{\rho}_1$, since the~$\rho_1$ itself does not
tend anywhere at all. One can only speak of the rate of convergence of the
{\it average values} of the dynamic quantities. If, for example, one takes
the characteristic function of certain region inside the vessel as a test
function~$\varphi$, then it will be just reasonable to consider the rate of equalization of the number of balls
in this  region.

To derive an expression for the limit distribution
(as~$N\rightarrow\infty$), we put~$3N=m+2$ and transform~(15)$\colon$
\begin{equation}
\overline{\rho}_1(u)=\frac{\Gamma\left(1+\frac{m}{2}\right)\intl_0^\infty{r}^{m-2}f\left(\frac{u_1^2+u_2^2+u_3^2+r^2}{2\varkappa}\right)dr}
{\pi^{3/2}\Gamma\left(1+\frac{m-3}{2}\right)\intl_0^\infty{r}^{m+1}f\left(\frac{r^2}{2\varkappa}\right)dr}.
\end{equation}
Here~$u=(u_1,u_2,u_3)$ and~$\Gamma$ is the gamma function. The density,
actually, depends on the velocity~$|u|$. Therefore, its~$k$-th moment
$$
\intl_{\mR^3}\bar{\rho}_1(|u|)|u|^k\,du_1\,du_2\,du_3
$$
is equal to
\begin{gather}
4\pi\intl_0^\infty\bar{\rho}_1(x)x^{k+2}\,dx=\notag\\
=\frac{2\Gamma\left(1+\frac{m}{2}\right)\Gamma\left(\frac{k+3}{2}\right)
\quad\intl_0^\infty\xi^{m+k+1}f\left(\frac{\xi^2}{2\varkappa}\right)\,d\xi}{\sqrt
{\pi}\,\Gamma\left(\frac{k+m}{2}+1\right)\quad\intl_0^\infty\xi^{m+1}f
\left(\frac{\xi^2}{2\varkappa}\right)\,d\xi}.
\end{gather}
When deriving this expression, we used~(16) and the elementary properties
of the gamma function.

Assume that the variable~$x$ takes all real values; then, it is natural
to consider the density~$\bar{\rho}_1(x)$ an even function. To simplify
the notation, we put~$\vk=1$ (or replace the function~$f(z)$
with~$f(\varkappa{z})$). Our goal is to show that,
as~$m\rightarrow\infty$, a distribution with the density
\begin{equation}
2\pi x^2\bar{\rho}_1(x),\,x\in\mR
\end{equation}
tends to the normal one.

To this end, we accept two assumptions$\colon$

\begin{itemize}
\item [(a)] the limit (as~$m\rightarrow\infty$) density~(18) has a finite
positive variance (the second moment,~$k=2$), and
\item [(b)] the function~$f$ has a summable derivative.
\end{itemize}

Condition~(a) is the condition of {\it non-degeneracy} of the limit
distribution, while condition~(b) is of technical nature and can probably
be weakened. Besides, $f$ should decay at infinity faster than any
 power function~$\colon$ otherwise the integrals in~(16) and~(17) are not
defined for every~$m$.

Putting~$k=2$ in~(17), integrating by parts and using assumption~(a), we obtain$\colon$
\begin{gather}
\lim_{m\rightarrow\infty}\intl_{-\infty}^\infty2\pi x^4\bar{\rho}_1(x)\,
dx=\notag\\
=\lim_{m\rightarrow\infty}\frac{3}{m+2}\frac{\intl_0^\infty\xi^{m+3}f\left
(\frac{\xi^2}{2}\right)\,d\xi}{\intl_0^\infty\xi^{m+1}f\left(\frac{\xi^2}
{2}\right)\,d\xi}=\notag\\
=-3\lim_{m\rightarrow\infty}\frac{\intl_0^\infty\xi^{m+3}f\left
(\frac{\xi^2}{2}\right)\,d\xi}{\intl_0^\infty\xi^{m+3}f'\left(\frac{\xi^2}
{2}\right)\,d\xi}=3c>0.
\end{gather}
Here~$3c$ is the variance, which exists due to~(a).

Let us put~$f+cf'=g$. Then, according to~(19),
\begin{equation}
\intl_0^\infty\xi^{m+3}g\Bigl(\frac{\xi^2}{2}\Bigr)\,d\xi\left/
\intl_0^\infty\xi^{m+3}f\Bigl(\frac{\xi^2}{2}\Bigr)\,d\xi\rightarrow0
\right.
\end{equation}
as~$m\rightarrow\infty$.

Now, we calculate the limit of the fourth moments~($k=4$):
\begin{equation}
\lim_{m\rightarrow\infty}\frac{2\Gamma\left(1+\frac{m}{2}\right)\Gamma
\left(\frac{7}{2}\right)}{\sqrt{\pi}\,\Gamma\left(\frac{m}{2}+3\right)}
\quad\frac{\intl_0^\infty\xi^{m+5}f\Bigl(\frac{\xi^2}{2}\Bigr)\,d\xi}
{\intl_0^\infty\xi^{m+1}f\Bigl(\frac{\xi^2}{2}\Bigr)\,d\xi}.
\end{equation}
According to~(20), the integral in the numerator of~(21) can be replaced
with the integral
$$
-c\intl_0^\infty\xi^{m+5}f'\left(\frac{\xi^2}{2}\right)\,d\xi=c(m+4)
\intl_0^\infty\xi^{m+3}f\left(\frac{\xi^2}{2}\right)\,d\xi.
$$
Using~(19), it is easy to calculate the limit~(21). It is equal
to~$1\cdot3\cdot5c^5$.

In the similar way, we prove that when~$k=2n$, the limit~(17),
as~$m\rightarrow \infty$, is
\begin{equation}
(2n+1)!!c^n.
\end{equation}
All the odd moments are, obviously, equal to zero.

Now, let~$\hat{\rho}_1$ be the density of the normal distribution in the
three-dimensional Euclidean space$\colon$
$$
\frac{1}{(\sqrt{2\pi}\sg)^3} e^{-\frac{u_1^2+u_2^2+u_3^2}{2\sg}}.
$$
Then
$$
2\pi x^2\hat{\rho}_1(x)=
\frac{1}{\sqrt{2\pi}\sg^3}x^2e^{-\frac{x^2}{2\sg}}.
$$
Let us calculate the variance of this distribution$\colon$
$$
\intl_{-\infty}^\infty\frac{x^4}{\sqrt{2\pi}
\sg^3}e^{-\frac{x^2}{2\sg}}\,dx=3\sg^2.
$$
The fourth moment is equal to~$1\cdot3\cdot5\sg^4$; more generally,
the~$2n$-th moment is equal to
\begin{equation}
(2n+1)!!\sg^{2n}.
\end{equation}

Formulas~(22) and~(23) coincide if we put~$c=\sg^2$. Hence, according to
the Chebyshev--Markov moment theorem (see Ref.~[11]),
\begin{equation}
\lim_{m\rightarrow\infty}\bar{\rho}_1(u)=\hat{\rho}(u),\,\,u\in\mR^3;\,\,
\sg=\sqrt{c}
\end{equation}

Up to now, we have been using the assumption that the function~$f$ does not depend on the
number of particles. In general, of course, this is not the case, and
instead of a single function~$f$, we have a  sequence of
functions,~$f_{m+2}$. Nevertheless, we can again
put~$g_{m+2}=f_{m+2}+cf'_{m+2}$ (provided that the limit distribution has a finite positive variance). It is easy to show that~(24) remains
valid if the limit expression~(20) is replaced with a more general
one$\colon$
$$
\intl_0^\infty\xi^{m+2k
+1}g_{m+2}\Bigl(\frac{\xi^2}{2}\Bigr)\,d\xi\left/\intl_0^\infty\xi^{m+2k+1}
f_{m+2}\Bigl(\frac{\xi^2}{2}\Bigr)\,d\xi\rightarrow0\right.
$$
as~$m\rightarrow\infty$ for any integer~$k\ge1$.

Recall that, for a fixed value of the total energy, the
distribution~(14) is called {\it microcanonical\/}. According to Maxwell
and Borel, it can be transformed to the {\it canonical} Gibbs distribution
if one assumes that the total energy is equal to~$NE$, the average
energy~$E$ of a single particle being independent of~$N$ (see Ref.~[12]).
Of course, this is an important assumption. A more general implementation
of this idea, applied to an ensemble of weakly interacting identical
subsystems, can be found, for example, in Ref.~[13]. In essence, the
author specifies the conditions, under which the microcanonical
distribution {\it weakly} converges to the canonical distribution as the
number of subsystems increases indefinitely. Besides, as test functions,
the author uses so called \emph{adders\/}, symmetrical functions of some specially chosen canonical variables. Our construction of the normal distribution is based on different ideas.

One should bear in mind that Boltzmann's and Bogolyubov's theories are not
free of all these problems, either. Suppose that, at the initial
time~$t=0$, the distribution~$\rho_N$ coincides with the
distribution~(14). The distribution~(14) is stationary, and corresponds to
the state of thermodynamical equilibrium (in Gibbs' approach). In
Boltzmann's theory, however, density~$\overline{\rho}_1(u)$ (which is
given by~(15)) corresponds, in the general case, to the initial
nonstationary distribution and should tend, in the course of time, to the
Maxwell distribution. In Bogolyubov's theory, we have a similar case$\colon$
 not every  summable function can be readily used as the density of the
initial distribution. It was supposed that the velocities in a particle
system should in some remote past (when the particles were far from each
other) be independent (so that every~$s$-particle distribution function
was reduced to a product of one-particle functions). It should be
underlined that this remote \emph{past} cannot be replaced with
the remote {\it future} (for discussion, see~[8]). However, in our
approach, the tendency to statistical equilibrium is invariant under the
time reversal.

\medskip

The work was supported by the Russian Foundation for Basic Research
(grant~01-01-22004) and the Foundation for Leading Scientific Schools
(grant~136.2003.1).


\vspace{2mm}


\begin{thebibliography}{99}

\bibitem{[1]}
    \author{V.\,V.\,Kozlov, D.\,V.\,Treshchev}
    \title{Weak convergence of solutions of the Liouville equation for nonlinear
    Hamiltonian systems}
    \journal*{Teor. Mat. Fiz.}
    \year{2003}
    \volume{134}
    \no{3}
    \page{388--400}
    \journal*{English transl.: Theor. Math. Phys.}
    \year{2003}
    \volume{134}
    \no{3}
    \page{339--350}

\bibitem{[2]}
    \author{V.\,V.\,Kozlov, D.\,V.\,Treshchev}
    \title{Evolution of measures in the phase space of nonlinear Hamiltonian systems}
    \journal*{Teor. Mat. Fiz.}
    \year{2003}
    \volume{136}
    \no{3}
    \page{496--506}
    \journal*{English transl.: Theor. Math. Phys.}
    \year{2003}
    \volume{136}
    \no{3}
    \page{1325--1335}

\bibitem{[3]}
    \author{V.\,V.\,Kozlov}
    \title{Billiards, Invariant Measures, and Equilibrium Thermodynamics}
    \journal*{Reg. \& Chaot. Dyn.}
    \year{2000}
    \volume{5}
    \no{2}
    \page{129--138}

\bibitem{[4]}
    \author{V.\,V.\,Kozlov}
    \title{Thermodynamics of Hamiltonian Systems and Gibbs
    Distribution}
    \journal{Dokl. Akad. Nauk}
    \year{2000}
    \volume{370}
    \no{3}
    \page{325--327}
    \journal{English transl.: Doklady Mathematics}
    \year{2000}
    \volume{61}
    \no{1}
    \page{123--125}

\bibitem{[5]}
    \author{V.\,V.\,Kozlov}
    \title{Kinetics of Collisionless Continuous Medium}
    \journal*{Reg. \& Chaot. Dyn.}
    \year{2001}
    \volume{6}
    \no{3}
    \page{235--251}

\bibitem{[6]}
    \author{Y.\,G.\,Sinai}
    \title{Dynamical Systems with Elastic Reflections. Ergodic Properties of
    Dispersing Billiards}
    \journal{Usp. Mat. Nauk}
    \year{1970}
    \volume{125}
    \no{2}
    \page{141--192}
    \journal*{English transl.: Russ. Math. Surv.}
    \year{1970}
    \volume{25}
    \page{137--189}

\bibitem{[7]}
    \author{V.\,V.\,Ten}
    \title{On normal Distribution in Velocities}
    \journal*{Reg. \& Chaot. Dynamics}
    \year{2002}
    \volume{7}
    \no{1}
    \page{11--20}

\bibitem{[8]}
    \author{G.\,E.\,Uhlenbeck, G.\,W.\,Ford}
    \title{Lectures in Statistical Mechanics}
    \journal*{Amer. Math. Soc.}
    \publisher{Providence}
    \year{1963}

\bibitem{[9]}
    \author{E.\,Fermi}
    \title{Thermodynamics}
    \publisher{New York: Prentice-Hall}
    \year{1937}

\bibitem{[10]}
    \author{M.\,A.\,Leontovich}
    \title{Introduction into Thermodynamics}
    \publisher{Moscow-Leningrad: Gostekhizdat}
    \year{1952}
    \translation{In Russian}

\bibitem{[11]}
    \author{N.\,I.\,Akhiezer}
    \title{The Classical Moment Problem}
    \publisher{Moscow: Nauka}
    \year{1961}
    \translation{In Russian}

\bibitem{[12]}
    \author{M.\,Kac}
    \title{Probability and Related Topics in Physical Sciences}
    \publisher{Interscience Publichers}
    \year{1958}

\bibitem{[13]}
    \author{F.\,A.\,Berezin}
    \title{Lectures on Statistical Physics}
    \publisher{Moscow-Izhevsk: Institute of Computer Science}
    \year{2002}
    \translation{In Russian}

\end{thebibliography}
\end{document}